\documentclass[a4paper,11pt]{article}
\pdfoutput=1 

\usepackage{jheppub} 

\usepackage[T1]{fontenc} 

\usepackage{graphicx} 

\usepackage{booktabs} 
\usepackage{array} 
\usepackage{paralist} 
\usepackage{verbatim} 
\usepackage{subfig} 

\usepackage{fancyhdr} 
\usepackage{sectsty}
\allsectionsfont{\sffamily\mdseries\upshape} 

\usepackage[nottoc,notlof,notlot]{tocbibind} 
\usepackage[titles,subfigure]{tocloft} 

\usepackage{amsfonts}

\usepackage{amsmath}
\usepackage{amssymb}
\DeclareMathOperator{\tr}{tr}
\DeclareMathOperator{\Tr}{Tr}

\title{\boldmath Emergent Supersymmetry in the Marginal Deformations of $\mathcal{N}=4$ SYM}

\author[a]{Qingjun Jin}

\affiliation[a]{Zhejiang Institute of Modern Physics,
Zhejiang University, 
Hangzhou, 310027, P. R. China \\ }

\emailAdd{qingjun@zju.edu.cn}

\abstract{We study the one loop renormalization group flow of the marginal deformations of $\mathcal{N}=4$ SYM theory using the $a$-function. We found that in the planar limit some non-supersymmetric deformations flow to the supersymmetric infrared fixed points described by the Leigh-Strassler theory. This means supersymmetry emerges as a result of renormalization group flow.}

\begin{document} 
\maketitle
\flushbottom

\section{Introduction}
\label{section1}

If the renormalization group(RG) flow of a quantum field theory has a stable conformal fixed point, and the fixed point preserves more symmetries(besides conformal symmetry itself) than the original theory, then these extra symmetries will emerge as a result of RG flow. Emergent supersymmetry has been found in different contexts, for example, in topological superconductors \cite{2014Sci...344..280G}, in gauge theories with some Yukawa operators \cite{Antipin:2011ny}, and in a class of models in 1+1 dimensions  \cite{Huijse:2014ata}. In this paper, we show that supersymmetry emerges in a four dimensional renormalizable quantum field theory: the marginal deformations of $\mathcal{N}=4$ SYM in the planar limit.

The most general superconformal deformation of $\mathcal{N}=4$ SYM is the Leigh-Strassler deformation \cite{Leigh:1995ep}. As a conformal field theory, the Leigh-Strassler deformation is a fixed subspace in the space of more general deformations. However, it is technically difficult to determine whether this fixed subspace is stable, given the huge number of parameters in the Yukawa and quartic scalar couplings.

The a-function (see e.g. \cite{Wallace:1974dx, Freedman:1998rd}) is a proposed quantity which increase monotonically with the energy scale, and its gradient flow gives the $\beta$-functions. The $a$-function can be very helpful in the study of RG flow in theories with large number of parameters, because the complicated behavior of RG flow in a high dimensional space is characterized by a single function.

We shall start in Section \ref{section2} by briefly reviewing the $a$-function, and discuss the relation between $a$-function and conformal fixed points. With the help of $a$-functions we study the flow of gauge and Yukawa couplings in gauge theories in Section \ref{section3}, and show that when the number of fermions and scalars satisfy a relation, the Yukawa couplings always flow to conformal fixed points. In Section  \ref{section4}, we briefly review the Leigh-Strassler theory.  And in Section  \ref{section5} and Section  \ref{section6}, we study the RG flow around the Leigh-Strassler theory, show that although the Leigh-Strassler theory seems to be a saddle point of generic deformations, it becomes stable if only a subspace of (but still non-supersymmetric) deformations are turned on.

\section{The $a$-function and the conformal fixed point}
\label{section2}

In two dimensional quantum field theories the Zamolodchikov c-theorem \cite{Zamolodchikov:1986gt, Zamolodchikov:1987ti}  identifies a C-function which satisfies a RG flow equation of the form
\begin{equation}\label{cfunction}
\frac{d C}{d\ln \mu}=\frac{3}{2}G_{IJ}\beta^I\beta^J,
\end{equation}
where $g^I$ are the couplings corresponding to the operators $O_I$, $\beta^I$ are beta functions of $g^I$. $G_{IJ}$ is proportional to the two point functions $\langle O_IO_J \rangle$, and it is positive definite, so \eqref{cfunction} implies that the C function increase monotonically with the energy scale. The four dimensional analog of C-theorem is the a-theorem, which conjectures that for four dimensional quantum field theories we can define the $a$-function, $\tilde{A}$, which satisfies
\begin{equation}\label{atg}
\partial_I\tilde{A}=T_{IJ}\beta^J,\ \frac{d\tilde{A}}{d\ln\mu}=G_{IJ}\beta^I\beta^J,
\end{equation}
where $G_{IJ}$ is the symmetric part of $T_{IJ}$, and a-theorem holds as long as $G_{IJ}$ is positive definite. The first evidence of $a$-theorem appeared in the 1970s \cite{Wallace:1974dx}, and a lot of progresses have been made in this direction since then (see \cite{ Kutasov:2003ux, Babington:2005vu,Komargodski:2011vj,Komargodski:2011xv, Fortin:2012cq, Jack:2013sha, Ramos:2014kla} and references therein).

\begin{figure}[htb]
\centering
\includegraphics[scale=0.5]{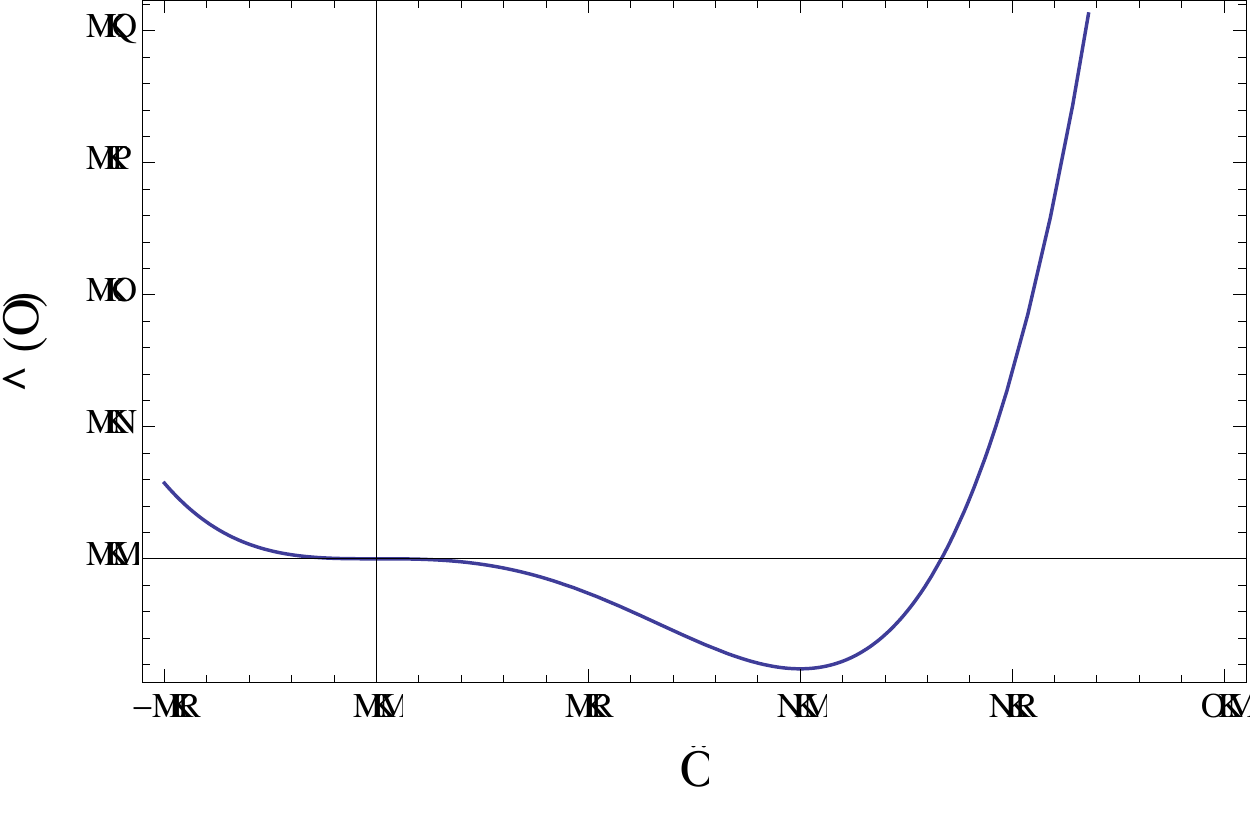}
\caption{$a$-function with $\tilde{A}(g)=\frac{1}{4}g^4-\frac{1}{3}g^3$. }
\label{fig:stable-nonstable}
\end{figure}

 In this paper we are mainly interested in the behavior of RG flow in the infrared region, and throughout the paper conformal fixed points always means infrared conformal fixed points unless otherwise specified. Conformal fixed points correspond to the stationary points of the $a$-function.  This is clear from \eqref{atg}: since $G_{IJ}$ is positive definite, $\frac{d\tilde{A}}{d\ln\mu}=0$ if and only if $\beta^I=0$.  A conformal fixed point can be isolated, in the sense that it is locally the only conformal fixed point, or it can be located in  a $D_c$ dimensional space of conformal fixed points.  An isolated conformal fixed point can be stable or unstable: a stable conformal fixed point corresponds to a local minimum of the $a$-function, while a unstable one corresponds to a local maximum or saddle points. An example of stable and unstable conformal fixed point is shown in Figure \ref{fig:stable-nonstable}: $g=1$ is a stable conformal fixed point, while $g=0$ is unstable.

The definition for a stable conformal fixed point $g^I_0$ is, any $g^I=g^I_0+\eta^I$ will flow to $g^I_0$ when energy goes to zero, as long as $\eta^I$ is small enough. But this is not true if $g^I_0$ located in  a $D_c$ dimensional space of conformal fixed points. So in this case it is more appropriate to discuss  the stableness of the conformal fixed subspace. A conformal fixed subspace, $\mathcal{C}$, is called stable, if any $g$ flow to a point in $\mathcal{C}$ at low energies, as long as $g$ is close enough to $\mathcal{C}$.

Stable fixed points(spaces) can be found by solving the differential equations numerically. The set of $g^I$ which flow to the given stable fixed points(spaces) has non-zero measure. So start with random $g^I$, when energy goes to zero, the possibility that the flow reaches the stable fixed point(spaces) is not zero.




A particularly interesting case of stable fixed points is when the $a$-function is bounded below: the RG flow must stop either when it reaches the lower bound, or trapped in some fixed point above the lower bound. This means the corresponding quantum field theory must flow to a conformal fixed point at low energies. In section \ref{section3} we will see the the first two orders of $\tilde{A}$ for gauge theories can be bounded blow with proper choice of fermion and scalar numbers.

\section{The RG flow of gauge and Yukawa couplings}
\label{section3}

In \cite{Jack:2014pua}, gauge theories with a Yukawa interaction $\frac{1}{2}\psi^T_iC(Y_a)_{ij}\psi_j\phi_a+h.c.$ and a quartic scalar interaction $\frac{1}{4!}\lambda_{abcd}\phi_a\phi_b\phi_c\phi_d$ was studied, and the $a$-function is computed to 4 loops, up to some $g^6$ terms\footnote{In \cite{Jack:2014pua} the coefficient of last term of \eqref{betas} was $-4$, we modified it to $-2$ based on our own calculations, and the results of other authors, e.g.  \cite{Luo:2002ti}.},
\begin{equation}
\begin{aligned}
ds^2=&G_{IJ}dg^Idg^J=\frac{2n_V}{g^2}(1+\sigma g^2)dg^2+\frac{1}{6}\tr[d\hat{y}_ady_a]+\frac{1}{144}d\lambda_{abcd}d\lambda_{abcd},\\
\tilde{A}^{(2)}=&-n_V\beta_0g^2,\\
\tilde{A}^{(3)}=&-\frac{1}{2}n_Vg^4(\beta_1+\sigma\beta_0)-\frac{1}{2}g^2 \tr[y_a\hat{y}_a\hat{C}^{\psi}] \\
&+\frac{1}{24}\tr[y_a\hat{y}_a y_b\hat{y}_b]
+\frac{1}{12}\Bigl(\tr[y_a\hat{y}_b y_a\hat{y}_b]+\frac{1}{4}\tr[y_a\hat{y}_b]\tr[ y_a\hat{y}_b]\Bigr),\\
\tilde{A}_{\lambda}^{(4)}=&\frac{1}{8}\lambda_{abcd}\lambda_{abef}\lambda_{cdef}
+\left(\frac{3}{2}g^4(t^{\phi}_At^{\phi}_B)_{ab}(t^{\phi}_At^{\phi}_B)_{cd}
-\frac{1}{2}\tr[y_a\hat{y}_by_c\hat{y}_d]\right)\lambda_{abcd}\\
&+\frac{1}{12}\lambda_{abcd}\lambda_{abce}\left(\tr[y_e\hat{y}_d]-6g^2(C^{\phi})_{ed}\right),\\
\end{aligned}\label{atilde}
\end{equation}
in which 
\begin{equation}
\begin{aligned}
\sigma=&\frac{1}{6}(102C_G-20R^{\psi}-7R^{\phi}),\ 
\beta_0=\frac{1}{3}\left(11C_G-2R^{\psi}-\frac{1}{2}R^{\phi}\right),\ \\
\beta_1=&\frac{1}{3}C_G\left(34C_G-10R^{\psi}-R^{\phi}\right)-\frac{1}{n_V}\tr[(C^{\psi})^2]
-\frac{2}{n_V}\tr[(C^{\phi})^2].\\
\end{aligned}\label{betas}
\end{equation}
$\tilde{A}_{\lambda}^{(4)}$ are the $\lambda$-dependent terms in $\tilde{A}^{(4)}$, we did not present the other terms here because we will not need them in this work. Our conventions for the group invariants and various other constants completely follows \cite{Jack:2014pua}, and for compactness we will not list them here.


$\tilde{A}^{(2)}$ only depend on $g$, and if $\beta_0<0$, $g=0$ is a the minimum of $\tilde{A}^{(2)}$, so at low energies, the gauge coupling approaches $0$, and gauge field decouples. Higher loop corrections do not affect our conclusion because they become unimportant when $g\rightarrow 0$.

When $\beta_0>0$, $g^2$ becomes greater at low energies(asymptotic freedom), and higher loop corrections becomes important. In order to find the minimum of $\tilde{A}^{(2)}+\tilde{A}^{(3)}$, we define the following quantities,
\begin{equation}
\begin{aligned}
&F_2=y_a\hat{y}_a-6g^2\hat{C}^{\psi},\\
&\mathcal{I}_{ab}=-i\tr(Y_a\bar{Y}_b)+i\tr(Y_b\bar{Y}_a),\\
&t_{ijkl}=Y_{aij}Y_{akl}+Y_{aik}Y_{ajl}+Y_{ail}Y_{ajk}.\\
\end{aligned}
\end{equation}
Using
\begin{equation}
\begin{aligned}
\frac{1}{3}t_{ijkl}\bar{t}^{ijkl}=&\tr(y_{a}\hat{y}_by_{a}\hat{y}_{b})
+\frac{1}{4}\tr(y_a\hat{y}_b)\tr(y_a\hat{y}_b)
-\frac{1}{4}\mathcal{I}_{ab}\mathcal{I}_{ab},\\
\end{aligned}
\end{equation}
the Yukawa terms in $\tilde{A}^{(3)}$ can be written as a sum of perfect squares,
\begin{equation}
\begin{aligned}
\tilde{A}^{(3)}=&-\frac{1}{2}n_Vg^4\left(b_1+\sigma\beta_0\right) 
+\frac{1}{24}\tr[F_2F_2]
+\frac{1}{36}t_{ijkl}\bar{t}^{ijkl}+\frac{1}{48}\mathcal{I}_{ab}\mathcal{I}_{ab},\\
\end{aligned}\label{a3new}
\end{equation}
where
\begin{equation}
b_1=\beta_1+\frac{3}{n_V}\tr[(\hat{C}^{\psi})^2]
=\frac{1}{3}C_G\left(34C_G-10R^{\psi}-R^{\phi}\right)+\frac{2}{n_V}\tr[(C^{\psi})^2]
-\frac{2}{n_V}\tr[(C^{\phi})^2].
\end{equation}

If $\beta_0>0$ and $b_1<0$, $\tilde{A}^{(2)}+\tilde{A}^{(3)}$ has a local minimum at\footnote{The stationary point of $\tilde{A}$ is given by $\nabla_I\tilde{A}=0$ instead of $\partial_I\tilde{A}=0$.}
\begin{equation}
F_2=\mathcal{I}_{ab}=t_{ijkl}=0,\ g^2=g^2_{m}\equiv-\frac{\beta_0}{b_1}.
\end{equation}

Notice $t_{ijkl}$ is proportional to the tree amplitude of 4 positive fermion,
\begin{equation}
A\left(\psi_i(k_1),\psi_j(k_2),\psi_k(k_3),\psi_l(k_4)\right)=\frac{[12][34]}{s_{12}}t_{ijkl}.
\end{equation}
and the vanishing of $t_{ijkl}$ forbids these UHV(ultra-helicity-violating) amplitude\footnote{It can be easily checked that $t_{ijkl}$ also vanishes for supersymmetric Yang-Mills theories.} at tree level. Actually we found the amplitude also vanishes at one loop if $t_{ijkl}=0$, and it is natural to expect it to vanish at all loops in conformal deformations of $\mathcal{N}=4$ SYM.

Now consider a theory with $n_{\psi}$ fermions and $n_{\phi}$ scalars, both in adjoint representation,
\begin{equation}
\begin{aligned}
&\beta_0=\frac{1}{3}C_G(11-2n_{\psi}-\frac{1}{2}n_{\phi}),\\
&\beta_1=\frac{1}{3}C_G^2\left(34-16n_{\psi}-7n_{\phi}\right),\\
&\sigma=\frac{1}{6}C_G(102-20n_{\psi}-7n_{\phi}).\\
\end{aligned}\label{beta01sigma}
\end{equation}

The pairs of $(n_{\psi},n_{\phi})$ satisfying $\beta_0>0,\ b_1<0$, and the corresponding $g^2_{m}N$ are collected in Table \ref{table1}. Notice for some choices  of $(n_{\psi},n_{\phi})$, $g^2_m N$ is still much smaller than 1,  and perturbation theory may be trusted.
\begin{table}
\begin{tabular}{|c|c|c|c|c|c|c|c|c|}
\hline
$(n_{\psi},n_{\phi})$&(1,6)&(1,7)&(1,8)&(1,9)&(1,10)&(1,11)&(1,12)&(1,13)\\
\hline
$g^2_m N$&$1$&$\frac{11}{26}$&$\frac{1}{4}$&$\frac{1}{6}$&$\frac{2}{17}$&$\frac{7}{82}$&$\frac{1}{16}$&$\frac{1}{22}$\\
\hline
$(n_{\psi},n_{\phi})$&(1,14)&(1,15)&(1,16)&(1,17)&(2,6)&(2,7)&(2,8)&(2,9)\\
\hline
$g^2_m N$&$\frac{1}{31}$&$\frac{1}{46}$&$\frac{1}{76}$&$\frac{1}{166}$&$1$&$\frac{7}{22}$&$\frac{1}{6}$&$\frac{1}{10}$\\
\hline
$(n_{\psi},n_{\phi})$&(2,10)&(2,11)&(2,12)&(2,13)&(3,6)&(3,7)&(3,8)&(3,9)\\
\hline
$g^2_m N$&$\frac{1}{16}$&$\frac{1}{26}$&$\frac{1}{46}$&$\frac{1}{106}$&$1$&$\frac{1}{6}$&$\frac{1}{16}$&$\frac{1}{46}$\\
\hline
\end{tabular}
\caption{$(n_{\psi},n_{\phi})$ satisfying $\beta_0>0,\ b_1<0$, and the corresponding $g^2_{m}N$.}\label{table1}
\end{table}

\section{The Leigh-Strassler Theory}
\label{section4}

The Leigh-Strassler theory is a $\mathcal{N}=1$ superconformal deformation of the $\mathcal{N}=4$ SYM. Superconformal symmetry and unitarity gives strong constraints to the anomalous dimensions of operators in supersymmetric theories, and a classification of supersymmetric deformations has been carried out in \cite{Cordova:2016xhm}. For $\mathcal{N}=1$ SYM it has been proved non-perturbatively that the conformal fixed point is always stable by showing there is a positive $a$-function around the fixed point \cite{Green:2010da}. The $a$-function was also computed  perturbatively in e.g. \cite{Anselmi:1997am,Anselmi:1997ys,Kutasov:2003ux,Barnes:2004jj}.

Less effort has been paid on the non-supersymmetric deformations of conformal field theories. The non-supersymmetric theories have much more parameters than the supersymmetric theories, which makes a direct study of RG flow unfeasible. And without supersymmetry, the known unitarity bounds are not enough to decide whether the deformation operators are relevant or irrelevant.

We will study the RG flow of non-supersymmetric theories using the $a$-function. From the perspective of last section, the Leigh-Strassler theory is a gauge theory with four chiral fermions six real scalars in adjoint representation of the $SU(N)$ gauge group.  Interestingly, from \eqref{beta01sigma} we find $\beta_0=\sigma=b_1=0$, and 
\begin{equation}
\begin{aligned}
\tilde{A}^{(2)}+\tilde{A}^{(3)}=&\frac{1}{24}\tr[F_2F_2]
+\frac{1}{36}t_{ijkl}\bar{t}^{ijkl}+\frac{1}{48}\mathcal{I}_{ab}\mathcal{I}_{ab}\ge 0.\\
\end{aligned}\label{a3ls}
\end{equation}
So $\tilde{A}^{(2)}+\tilde{A}^{(3)}$ has a global minimum at
\begin{equation}\label{conditionls}
F_2=\mathcal{I}_{ab}=t_{ijkl}=0.
\end{equation}
Eq. \eqref{conditionls} is 'homogeneous' in $g$ and $Y_{aij}$: if $g$ and $Y_{aij}$ solves \eqref{conditionls}, $|z| g$ and $z Y_{aij}$ also solves \eqref{conditionls} for arbitrary non-zero complex $z$. This implies unlike the theories in Table \ref{table1}, where $g^2_mN$ is fixed, conformal fixed points may exist for arbitrary gauge couplings when $(n_{\psi},n_{\phi})=(4,6)$.

We will focus on the planar limit, then in terms of $SU(N)$ matrix-valued fields, the Yukawa interaction can be written as $Y_{IAB}\Tr(\phi^I\psi^A\psi^B)+\bar{Y}^{IBA}\Tr(\phi^I\bar{\psi}_A\bar{\psi}_B)$ and the quartic scalar coupling is $\frac{1}{4}\lambda_{IJKL}\Tr(\phi^I\phi^J\phi^K\phi^L)$.

It is convenient to combine $\phi^I$ into 3 complex scalars $\phi^i$, and to discriminate $\psi^i$ and $\psi^4$. $\psi^i$ are the super-partner of $\phi^i$, while $\psi^4$ is the super-partner of the gauge field.  The Yukawa interaction of  the Leigh-Strassler theory is given by
\begin{equation}
\begin{aligned}
L_Y=&\Tr\left[\kappa \epsilon^+_{ijk}\phi^i\left(q\psi^j\psi^k- \frac{1}{q}\psi^k\psi^j\right)
+h\sum_i\phi^i\psi^i\psi^i+g\bar{\phi}_i[\psi^i,\psi^4]\right]+c.c.
\end{aligned}\label{lyls}
\end{equation}
where
\begin{equation}
\epsilon^+_{ijk}=\frac{1}{2}(\epsilon_{ijk}+|\epsilon_{ijk}|),
\end{equation}
and $h$ and $q$ are two complex parameters.

The quartic scalar interactions are related to the Yukawa couplings $Y_{ijk}$ by,
\begin{equation}\label{lphinonplanar}
L_{\phi}=-\frac{g^2}{2}\Tr([\phi^i,\bar{\phi}_i]^2)-Y_{ijm}\bar{Y}^{lkm}\Tr(\phi^i\phi^jT^a)\Tr(\bar{\phi}_k\bar{\phi}_lT^a),
\end{equation}
where $T^a$ are the $SU(N)$ generators.

Besides supersymmetry, the theory is also invariant under a $U(1)$ transformation:
\begin{equation}
\psi^i\rightarrow e^{i\xi}\psi^i,\ \psi^4\rightarrow e^{-3i\xi}\psi^i,\ \phi^i\rightarrow e^{-2i\xi}\phi^i.
\end{equation}

In order for the theory to be conformal, $\kappa$, $h$ and $q$ must satisfy a condition. At 2 loops and in the planar limit, this condition is
\begin{equation}
2g^2=|h|^2+\kappa^2(|q|^2+|q|^{-2}).
\end{equation}
The condition under which the theory is conformal up to three loops (four loops in planar limit) was given in \cite{Bork:2007bj}.

In the planar limit, \eqref{conditionls} becomes
\begin{equation}
\begin{aligned}
(F_2)_A^B=&Y_{IAC}\bar{Y}^{IBC}+Y_{ICA}\bar{Y}^{ICB}-6g^2\delta_A^B,\ \\
\mathcal{I}_{IJ}=&\frac{1}{i}(Y_{JCD}\bar{Y}^{ICD}-Y_{ICD}\bar{Y}^{JCD})=0,\\
t_{ABCD}=&Y_{IAB}Y_{ICD}+Y_{IDA}Y_{IBC}.\\
\end{aligned}\label{FIT}
\end{equation}
It can be verified that \eqref{FIT} holds\footnote{Actually \eqref{conditionls} hold even in the non-planar case for the Leigh-Strassler theory.} and the Leigh-Strassler theory does lie in the the global minimum of $\tilde{A}^{(2)}+\tilde{A}^{(3)}$. However, it is possible that there are other (non-supersymmetric) conformal fixed points in the neighborhood of the Leigh-Strassler deformation, then the RG flow may end up reaching these non-supersymmetric conformal fixed points, and supersymmetry fails to emerge. in order to exclude this possibility we will examine the anomalous dimensions of marginal operators in the next section.

\section{Marginal Operators and Conformal Deformations}
\label{section5}

In this section we study the conformal deformations of the Leigh-Strassler theory. The space of 'all conformal deformations' may have multiple components, and in different components, the space may have different dimensions. So to be more accurate we define the term 'sector':  the sector of a given conformal fixed point is the irreducible component containing a neighborhood of the point in the space of conformal deformations. The Leigh-Strassler deformation has 4 real physical parameters, so locally it is the only conformal fixed subspace if the enclosing sector also has 4 physical parameters.

Different fixed points in the same sector are physically equivalent if they are related by $SO(6)$ and $U(4)$ redefinitions of scalars and fermions. So number of 'physical' parameters is the dimension of the quotient space of the sector by $SO(6)\times U(4)$.


Suppose $L_{cft}$ is the Lagrangian of the enclosing sector of the Leigh-Strassler theory, and $g_a$ are the physical parameters,
\begin{equation}
\begin{aligned}
L_{cft}=&L_{cft}(g_a, A_{\mu}, U^A_B(\omega_i)\psi^A, O^{IJ}(\omega_i)\phi^J),\\
\end{aligned}
\end{equation}
where $\omega_i$'s are parameters of U(4) (16 parameters) and SO(6) (15 parameters) redefinitions $U^A_B$ and $O^{IJ}$.

General deformations can be written as $L_0$, the Lagrangian of the Leigh-Strassler theory, plus four types of  dimension-4 operators,
\begin{equation}
L_{\mathcal{O}}=L_{cft}^0+\delta g_a\frac{\partial L_{cft}}{\partial g_a}+\delta \omega_i\frac{\partial L_{cft}}{\partial \omega_i}+c^m O_m+z^{\alpha}Z_{\alpha}.
\end{equation}
The first type, $\frac{\partial L_{cft}}{\partial g_a}$, corresponds to the variation of the Lagrangian when the parameters $g_a$ changes. The number of $\frac{\partial L_{cft}}{\partial g_a}$ is $N_p$, which is the number of physical parameters The second type, $\frac{\partial L_{cft}}{\partial \omega_i}$, corresponds to the variation of the Lagrangian under the redefinition of scalars and fermions. The number of $\frac{\partial L_{cft}}{\partial \omega_i}$ is at most 31, but it is possible that a subgroup of $U(4)\times SO(6)$, $G_{sym}$, is preserved in the theory. In this case, the corresponding $\mathcal{O}_i$ vanishes, so the number of $\mathcal{O}_i$ is the same as the number of generators of the quotient group $U(4)\times SO(6)/G_{sym}$. For example, $\mathcal{N}=4$ SYM has a $SU(4)$ symmetry, so the number of $\mathcal{O}_i$ is only 16. The $\gamma_i$-deformed SYM \cite{Frolov:2005iq,Frolov:2005dj} has a $U(1)^3$ symmetry, and the number of $\mathcal{O}_i$ is 28. Adding these two types of operators to the Lagrangian does not break the conformal symmetry, and the corresponding beta functions vanishes.

The third type, $O_m$, are operators with non-zero anomalous dimensions, or marginally irrelevant operators. The last type, $Z_{\alpha}$ are operators which break conformal symmetry but with vanishing anomalous dimensions. We will call operators with zero anomalous dimensions protected operators, $Z_{\alpha}$ will be called accidentally protected operators, and the number of $Z_{\alpha}$ will be denoted by $N_{acci}$.

Each protected operator corresponds to a zero eigenvalue of the anomalous dimension matrix $\nabla_I\beta^J$, so we have
\begin{equation}\label{npex}
N_p=Dim(Ker(\nabla_I\beta^J))-N_{acci}-Dim(U(4)\times SO(6))+Dim(G_{sym}).
\end{equation}

As an example, we consider the Yukawa couplings of $\gamma_i$ deformed $\mathcal{N}=4$ SYM. There are in all 34 protected operators, in which none is accidentally protected, and the theory preserves a $U(1)^3$ symmetry. From \eqref{npex} the theory has 6 physical parameters. The complete Lagrangian of this 6-parameter theory is not known yet, but it has been formulated for a 3-parameter sub-theory(when all $\gamma_i$'s are real) \cite{Frolov:2005iq}. The gravity dual of this 3-parameter sub-theory \cite{Frolov:2005iq} is a sub-theory of a 6+2 parameter\footnote{These 2 extra parameters corresponds to the variation of complex gauge coupling $\tau=\frac{1}{g^2}+\frac{i\theta}{8\pi^2}$.} deformation of $AdS_5\times S^5$ \cite{Frolov:2005dj}. This indicates the $6+2$ parameter deformation is the gravity dual of the enclosing sector of $\gamma_i$ deformed SYM. 

The Leigh-Strassler deformation has 46 protected Yukawa operators \footnote{The anomalous dimensions of quartic scalar operators will be modified when these protected Yukawa operators are added to the Lagrangian. This is why in \eqref{accidental}, protected Yukawa operators also contain a quartic scalar piece.}. Among them 30 operators correspond to the variation of the Lagrangian under SO(6) and U(4) rotations of scalars and fermions. 4 protected operators corresponds to the variation of the Lagrangian when the parameter of the theory, $\beta$ and $h$ change. The left 12 operators are\footnote{In \eqref{accidental} $O_1^i$ and $O_2^i$ are both complex operators. Each of them corresponds to 2 real accidentally protected operator which contain both chiral and anti-chiral fermions.}:
\begin{equation}
\begin{aligned}
O_1^i=&\Tr\Bigl(\kappa(p^2+\frac{1}{p^2}-1)\phi^i\psi^4\psi^4
+\frac{\kappa}{\bar{h}}(\bar{q}-\bar{q}^{-1})^2(\phi^1)^4\\
&+(\phi^1)^2(\bar{q}\phi^2\phi^3-\bar{q}^{-1}\phi^3\phi^2)
+(\bar{q}-\bar{q}^{-1})\phi^1\phi^2\phi^1\phi^3-\frac{\bar{h}}{\kappa}(\phi^2\phi^3)^2\Bigr),\\
O_2^{i}=&\Tr\Bigl(\phi^j\{\psi^k,\psi^4\}+\phi^k\{\psi^j,\psi^4\}-\frac{\kappa}{\bar{h}}(\bar{q}-\bar{q}^{-1})\phi^i\{\psi^i,\psi^4\},\\
&+(\phi^j)^2[\bar{\phi}_j,\phi^k]+(\phi^k)^2[\bar{\phi}_k,\phi^j]
+[\phi^i,\bar{\phi}_i]\{\phi^j,\phi^k\}\\
&-\frac{\kappa}{\bar{h}}(\bar{q}-\bar{q}^{-1})(\phi^i)^2([\phi^j,\bar{\phi}_j]+[\phi^k,\bar{\phi}_k])\Bigr),\\
\end{aligned}\label{accidental}
\end{equation}
Adding these operators to the Leigh-Strassler Lagrangian,
\begin{equation}
\delta L=a_i O_1^i+\bar{a}_i \bar{O}_1^i+b_i O_2^i+\bar{b}_i \bar{O}_2^i,
\end{equation}
$t_{ijkl}$ and $\mathcal{I}_{ab}$ are invariant, but $F_2$ is not. For example,
\begin{equation}
\begin{aligned}
\frac{1}{2}\delta (F_2)^1_1=&\frac{\kappa^2|q-1/q|^2}{|h|^2}|b_1|^2+|b_2|^2+|b_3|^2,\\
\frac{1}{2}\delta (F_2)^4_4=&|a_1|^2+|a_2|^2+|a_3|^2+
\left(2-\frac{\kappa^2|q-1/q|^2}{|h|^2}\right)(|b_1|^2+|b_2|^2+|b_3|^2).
\\
\end{aligned}
\end{equation}
Apparently turning on any of these operators will change $F_2$, and in the end increase $\tilde{A}$. So these operators are accidentally protected operators, and do not corresponds to new conformal field theories.
From \eqref{npex}, the number of parameters in the enclosing sector is $N_p=4$, so we have proved that as far as Yukawa couplings are concerned, locally the Leigh-Strassler deformation is the only conformal deformation.

Last,  let us emphasize that these 12 accidentally protected operators may not be protected by higher loop corrections. If the two loop anomalous dimensions turn out to be negative, the Leigh-Strassler theory will be unstable even at weak coupling. However, it will be guaranteed  to be stable at weak couplings if we complete turn off the accidentally protected operators, for example, in the subspace of deformations
described by
\begin{equation}
\begin{aligned}
L_Y=&\Tr\left[Y_{ijk}\phi^i\psi^j\psi^k+X^i_j\bar{\phi}_i\psi^j\psi^4
+Z^i_j\bar{\phi}_i\psi^4\psi^j\right]+c.c.\\
L_{\phi}=&\lambda_{ij\bar{k}\bar{l}}\Tr(\phi^i\phi^j\bar{\phi}_k\bar{\phi}_l)
+\frac{1}{2}\lambda_{i\bar{k}j\bar{l}}\Tr(\phi^i\bar{\phi}_k\phi^j\bar{\phi}_l).\\
\end{aligned}\label{subspace}
\end{equation}

\section{Emergent Supersymmetry}
\label{section6}


In the planar limit, the $\lambda$-dependent terms of $\tilde{A}^{(4)}$ in \eqref{atilde} is reduced to
\begin{equation}
\begin{aligned}
\tilde{A}^{(4)}_{\lambda}=&-\frac{3}{8}g^4\lambda_{IIJJ}-\frac{3}{4}g^2\lambda_{IJKL}\lambda_{LKJI}+\frac{1}{8}D_{MN}\lambda_{MIJK}\lambda_{KJIN},\\
&+B_{IJKL}\lambda_{LKJI}-\frac{1}{6}\lambda_{IJKL}\lambda_{LKMN}\lambda_{NMJI},\\
\end{aligned}\label{scalarV}
\end{equation}
in which $D_{IJ}$ corresponds to a fermion bubble diagram,
\begin{equation}
D_{IJ}=Y_{ICD}\bar{Y}^{JCD}+Y_{JCD}\bar{Y}^{ICD},
\end{equation}
and $B_{IJKL}$ corresponds to a fermion box diagram,
\begin{equation}
B_{IJKL}=Y_{IAB}\bar{Y}_J^{AD}Y_{KCD}\bar{Y}_L^{CB}
+\bar{Y}_I^{AB}Y_{JCB}\bar{Y}_K^{CD}Y_{LAD}.
\end{equation}

$\tilde{A}^{(4)}_{\lambda}$ is a polynomial of degree 3 in $\lambda_{IJKL}$ so it does not have a global minimum. Nevertheless, $V_S$ may have local minimum or saddle points, corresponding to stable or unstable fixed points of single trace scalar couplings, respectively.

Numerical tests shows that the anomalous dimension matrix of $\lambda_{IJKL}$ is still positive semi-definite. There are 6 protected scalar operators, and they are combinations of three holomorphic and three anti-holomorphic operators.
\begin{equation}
\mathcal{O}^R_i=\frac{1}{2}\Tr(\mathcal{O}_i+\bar{\mathcal{O}}_i),
\ \mathcal{O}^I_i=\frac{1}{2i}\Tr(\mathcal{O}_i-\bar{\mathcal{O}}_i).
\end{equation}

 One anti-holomorphic operators is
\begin{equation}
\begin{aligned}
\bar{\mathcal{O}}_1=&\frac{\kappa}{h}\left[h^3(q^2+\frac{1}{q^2})+\kappa^3(q-\frac{1}{q})^3\right]\bar{\phi}_1^4
+h^2\kappa(q^2+\frac{1}{q^2}-1)\bar{\phi}_1(\bar{\phi}_2^3+\bar{\phi}_3^3)\\
&+\left[h^3\frac{1}{q}+\kappa^3(q^2-1)^2\right]\bar{\phi}_1^2\bar{\phi}_2\bar{\phi}_3
+\left[-h^3q+\kappa^3(\frac{1}{q^2}-1)^2\right]\bar{\phi}_1^2\bar{\phi}_3\bar{\phi}_2\\
&+(q-\frac{1}{q})\left[-h^3+\kappa^3(q-\frac{1}{q})\right]\bar{\phi}_1\bar{\phi}_2\bar{\phi}_1\bar{\phi}_3\\
&-h\kappa^2(q-\frac{1}{q})(q^2+\frac{1}{q^2}-1)\bar{\phi}_2^2\bar{\phi}_3^2
+\frac{h}{\kappa}\left[h^3-\kappa^3(q-\frac{1}{q})\right]\bar{\phi}_2\bar{\phi}_3\bar{\phi}_2\bar{\phi}_3.\\
\end{aligned}
\end{equation}

The other two anti-holomorphic operators can be obtained form $\bar{\mathcal{O}}_1$ using the $\mathbb{Z}_3$ symmetry. The holomorphic operators are the Hermitian conjugate of anti-holomorphic operators.

If we add these protected operators to the Leigh-Strassler Lagrangian,
\begin{equation}
L= L_{LS}+z_i\mathcal{O}_i+\bar{z}_i\bar{\mathcal{O}}_i.
\end{equation}
We can expand $V_S$ as the power of $z_i$ and $\bar{z}_i$. The $\mathcal{O}(z)$ order vanishes because it is  proportional to beta functions in Leigh-Strassler theory. The $\mathcal{O}(z^2)$ order also vanishes because it is proportional to the anomalous dimensions of accidentally protected operators $\mathcal{O}_i$ and $\bar{\mathcal{O}}_i$.
\begin{equation}\label{vschangebyz}
V_S(z_i,\bar{z}_i)=V_S^{LS}-\frac{1}{6}\delta \lambda_{IJKL}\delta \lambda_{LKMN}\delta \lambda_{NMJI},
\end{equation}
in which $\delta \lambda$ is defined by
\begin{equation}
\frac{1}{4}\delta \lambda_{IJKL}\Tr(\phi^I\phi^J\phi^K\phi^L)=z_i\mathcal{O}_i+\bar{z}_i\bar{\mathcal{O}}_i\ .
\end{equation}
With complex indices, the only non-vanishing components of $\delta \lambda$ are $\lambda_{ijkl}$ and $\lambda_{\bar{i}\bar{j}\bar{k}\bar{l}}$, so $\delta \lambda$'s cannot give non-zero contribution to $\lambda^3$ terms in eq. \eqref{vschangebyz}. So the presence of $\delta \lambda$ does not change $V_S$. But we found that $\delta \lambda$ make the beta functions non-zero,
\begin{equation}
\frac{dV_S(z_i,\bar{z}_i)}{d\ln\mu}=\mathcal{O}(z^2).
\end{equation}
One can add $\mathcal{O}(z^2)$ order operators to the Lagrangian to cancel these $\mathcal{O}(z^2)$, and $\mathcal{O}_i$ will become exact marginal if the same can be done to all orders in $z$. It is technically hard to exclude this possibility, but numerical tests indicates it fails at $\mathcal{O}(z^3)$. So the Leigh-Strassler deformation seems to be a saddle point in the complete parameter space.

In the subspace of deformations described by \eqref{subspace}, the operators $\mathcal{O}_i$ are turned off, and the Leigh-Strassler theory becomes stable. So $\mathcal{N}=1$ supersymmetry emerges at low energies in this non-supersymmetric subspace.


\section{Discussions}
\label{section7}

In this paper we focused on the flow of deformed $\mathcal{N}=4$ SYM in the planar limit. At the non-planar level, the Leigh-Strassler deformation is a saddle fixed point even in the subspace \eqref{subspace}. The next step along this route shall be finding the maximal subspace of deformations in which the Leigh-Strassler deformation is a stable fixed point even at non-planar level.

Besides supersymmetry, other symmetries may also emerge in other models. In fact, since the Leigh-Strassler deformation preserves a $U(1)$ symmetry while the subspace \eqref{subspace} does not, this $U(1)$ also emerges together with supersymmetry. It would be worthwhile to check whether the $\gamma_i$ deformed SYM, which preserves a $U(1)^3$ symmetry is a stable fixed point in some subspace of deformations.

In Section \ref{section3} we proved Yukawa couplings of deformed $\mathcal{N}=4$ SYM always flow to fixed points. Following the flow one may find new types of fixed points(subspaces) which is previously unknown.

Last but not least, it would be interesting to search for more realistic models in which supersymmetry emerges. In these models supersymmetry are not treated as a fundamental symmetry. Even if the 'fundamental theory' is non-supersymmetric, supersymmetry may still emerge as a result of RG flow at some intermediate energy scale which is much lower than the characteristic energy scale of the 'fundamental theory', but still much higher than the electroweak scale.

\appendix

\acknowledgments

It is a pleasure to thank Radu Roiban and Bo Feng for helpful physics discussions and comments on the manuscript. Most calculations in the project was finished in Pennsylvania State University, and was supported by the US DoE under contract DE-SC0008745. Part of the paper was finished in Zhejiang University, and was supported by Qiu-Shi funding and Chinese NSF funding under contracts No.11031005, No.11135006 and No.11125523.


\bibliographystyle{JHEP}
\bibliography{/Users/jin/Documents/tex/PSUThesis/Biblio-Database}{}

\end{document}